\documentclass[amsfonts, amsmath, amssymb,
 aip, jcp, reprint, showpacs, superscriptaddress,
 ]{revtex4-1}

\usepackage{graphicx}
\usepackage{dcolumn}
\usepackage{bm}
\usepackage{hyperref}
\usepackage{amsmath}

\usepackage{color}
\usepackage{soul}

\newcommand{\pfdep}{$\eta^{3/2}$}

\newcommand{\Tdep}{$T^{1/2}$}
\newcommand{\THdep}{$T^{1/4}$}

\newcommand{\msnvpffTdep}{$m^{-1/2}(N/V)^{2/3}\eta^{2/3}T^{0}$}
\newcommand{\msnvpfTdep}{$m^{-1/2}(N/V)^{2/3}\eta^{3/2}T^{1/4}$}

\begin{document}

\title{Thermal Conductivity of Simple Liquids:\\
Origin of Temperature and Packing Fraction Dependences }

\author{Yoshiki Ishii}
\author{Keisuke Sato}
\affiliation{Graduate School of Science and Technology, Niigata University,
 8050 Ikarashi 2-no cho, Nishi-ku, Niigata 950-2181, Japan}

\author{Mathieu Salanne}
\affiliation{Sorbonne Universit\'{e}s, UPMC Univ Paris 06, UMR 8234,
PHENIX, F-75005 Paris, France}
\affiliation{CNRS, UMR 8234, PHENIX, F-75005 Paris, France}

\author{Paul A. Madden}
\affiliation{Department of Materials,
University of Oxford, Parks Road, Oxford OX1 3PH, United Kingdom}

\author{Norikazu Ohtori}
 \email[]{ohtori@chem.sc.niigata-u.ac.jp}
\affiliation{Department of Chemistry, Niigata University, 8050 Ikarashi 2-no cho,
 Nishi-ku, Niigata 950-2181, Japan}
\date{\today}

\begin{abstract}
The origin of both weak temperature dependence and packing fraction dependence of \THdep\pfdep 
in the thermal conductivity of the simple Lennard-Jones (LJ) liquid is explored. 
In order to discuss the relative contributions from attractive or repulsive part of the interaction potential separately, 
the thermal conductivity of a series of Weeks-Chandler-Anderson (WCA) fluids is calculated by molecular dynamics simulations. 
The results show that the repulsive part plays the main role in the heat conduction, 
while the attractive part has no direct effect on the thermal conductivity for a given packing fraction. 
By investigating WCA fluids with potentials of varying softness, 
we explain the difference observed between the LJ liquids such as argon and Coulombic liquids such as NaCl.  
\end{abstract}

\maketitle

\section{\label{sec:level1} INTRODUCTION}
In studies of transport properties of liquids, 
molecular dynamics (MD) simulation are widely used as computational experiments. 
Not only are they able to predict the relevant physical properties, 
but they also provide useful reference data for testing theoretical approaches. 
Here we focus on the thermal conductivity, 
a quantity for which there is an increasing demand for reliable data in the framework of energy production and storage applications. 
For example, in concentrating solar power plants~\cite{Cartlidge2011} or
in prospective Generation IV nuclear reactors,~\cite{waldrop2012a} the heat transfer fluids are molten salts. 
In these liquid materials, experiments are often very difficult to perform, 
and MD has appeared as a viable way for filling the gap in the databases.~\cite{Ohtori2009:2,Salanne2011} 
It would nevertheless be very valuable to develop some reliable analytical expressions.

The most important efforts made in this direction are those of Heyes {\it et al}.
These authors have extensively investigated fluids with repulsive interaction potentials 
of different softness\cite{Heyes2001,Branka2004,Heyes2005} (but close to the hard sphere limit) 
and focused on several transport properties, including thermal conductivity. 
They have also studied carefully the packing fraction dependence. 

Here we will focus on fluids where the interaction potentials have a somewhat softer character. 
On the one hand,  
we have recently shown that the thermal conductivities of simple liquids described with the Lennard-Jones (LJ) interaction potentials 
follow a linear scaling law as a function of \msnvpfTdep, 
where $m$ is the atomic mass, $V$ the volume of the system, $N$ the number of atoms included in $V$, 
$\eta$ the packing fraction, and $T$ the absolute temperature.~\cite{Ohtori2014} 
This result was extracted from a series of MD simulation 
where we performed a systematic scaling of the particle sizes and/or of the volume of the simulation cell. 
On the other hand, in the case of simple molten salts such as alkali halides, 
the temperature and packing fraction dependence is much weaker than that of the LJ liquids 
and the thermal conductivity can simply be expressed as a function of \msnvpffTdep.~\cite{Ohtori2014,ohtori2009}
It is worth noting that in both cases the calculations results agree very well with the available experimental data. 
In the two proposed functions,  the packing fraction plays the most significant role, and there is small or null explicit temperature dependence. 

The objective of this work is to explore the microscopic origin of these observations. 
To this end, we have undertaken a series of MD simulation using Weeks-Chandler-Andersen (WCA)
interaction potentials~\cite{Weeks1971} of varying softness, from which we calculated
the thermal conductivity. The parameters of the interaction potentials were chosen 
in order to allow straightforward comparisons with the corresponding LJ fluid.
We then interpret the thermal conductivity data available for liquid NaCl in light of these results.

\section{\label{sec:level2} SIMULATIONS}
MD simulations were carried out with the WCA potential, in which the interaction between two atoms $i$ and $j$,
with positions ${\bf r}_i$ and ${\bf r}_j$ and separated by a distance $r_{ij}=\mid{\bf r}_i-{\bf r}_j\mid$ is defined as
     \begin{eqnarray}
      \label{eq:potential}
      {\phi} \left( r_{ij} \right) =  
        \begin{cases}
        \! 4 \epsilon \! \left[ \left( \dfrac{\sigma / a} {r_{ij}} \right)^{  \! \!  2n}   \! \!\! \!   -
                   \left( \dfrac{\sigma / a}{r_{ij}} \right)^{  \! \!  n} \right]  \! \! + \epsilon ,
                                  &    \! \! \! \!   r_{ij} \leq \sqrt[n]{2} \sigma /a  \\
        \ \ \ \ \ \ \ 0 , &    \! \! \! \!  r_{ij} > \sqrt[n]{2} \sigma /a,
        \end{cases}
     \end{eqnarray}
\noindent In this potential, the intermolecular forces are entirely repulsive, and when $n$ equals 6 
 they are identical to the repulsive forces in a LJ fluid using the same $\epsilon$ and $\sigma$ parameters. 
The latter
were taken from the potential given by Maitland~\cite{Maitland}
 ($\epsilon / k_{\rm B} = 119.8~{\rm K}$ and $\sigma = 0.3405~{\rm nm}$) 
to reproduce
thermodynamic properties of liquid argon.~\cite{Hoef1999}
It yields the correct phase diagram including the liquid-vapor coexistence curve.~\cite{Anta1997,Hoef1999}
Finally $a$ is a coefficient introduced in order to scale particle size $\sigma$.
It
allows us to vary the packing fraction $\eta=\pi(\sigma /a)^3N/6V$
at constant number density.~\cite{Ohtori2014}
The effect of the softness of the interaction potential was examined by using $n$~=~2, 3, 4, 6, 9, and 12: a smaller value corresponds to a softer potential. 
All the MD simulations were performed in the $NVE$ ensemble after careful equilibration
(the initial configuration was a $fcc$ crystalline structure).
The number of particles was fixed to 864 and the density was set at 1124.9~$\rm{kg m}^{-3}$, 
which is the experimental density of liquid argon~\cite{Gilgen1994} at a temperature of 124~K and a pressure of 2~MPa. 
We obtained an average pressure of 0.2~MPa (with a standard deviation of 4~MPa) for the simulation where the scaling coefficient $a$ was set to 1. Note that when we perform a $NPT$ simulation at the corresponding $T$ and $P$ conditions, the density is underestimated by 5~\%. In our previous work we have shown that a quantitative agreement with the experimental thermal conductivity was obtained when using the correct density, so that we have chosen to follow a similar setup here.~\cite{Ohtori2014}
A time step of 10~fs was used to integrate the equations of motion using the velocity Verlet algorithm.~\cite{Allen} 
During the equilibration, the kinetic energy was kept constant at each temperature 
by scaling the velocities of all atoms during the initial 10~ps, 
and then the calculation was carried out for another 190~ps without further scaling.  
Production runs of 2.5~ns were performed for the calculation of the thermal conductivity. 
The velocities of all the atoms were corrected every 5000~steps in order to keep the total momentum of the system negligible. 

For LJ liquids, the thermal conductivity is given by the Green-Kubo formula~\cite{Hansen} as
\begin{equation}
   \lambda = \dfrac{1}{3k_BVT^2} \int_0^{\infty}\left< {\bf J}^{e}(t){\bf J}^{e}(0)\right> {\rm d}t,
\end{equation}
where $k_B$ is the Boltzmann constant, and ${\bf J}^e$ the energy current, 
\begin{eqnarray}
{\bf J}^e(t)&=&\sum_{i=1}^N\left(\frac{1}{2}m_i\mid{\bf v}_i\mid^2 +\sum_{j\ne i}^N\frac{1}{2}\phi(r_{ij})\right){\bf v}_i \\
& & +\frac{1}{2}\sum_{i=1}^N\sum_{j\ne i}^N\left({\bf F}_{ij}\cdot {\bf v}_i \right) {\bf r}_{ij}. \nonumber
\end{eqnarray}
\noindent In this expression ${\bf v}_i$ is the velocity of atom $i$ and ${\bf F}_{ij}$ is the force deriving from the potential $\phi(r_{ij})$. 

\section{\label{sec:level3} RESULTS AND DISCUSSION}

\begin{figure}[t]
\includegraphics[width=\columnwidth]{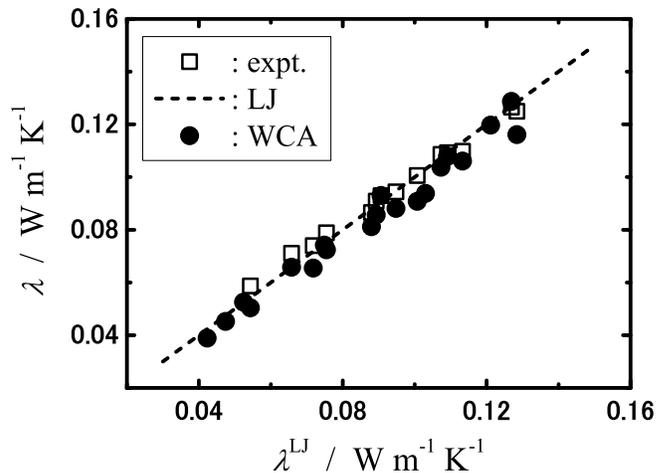}
\caption{\label{WCA} Comparison of the thermal conductivity calculated with the WCA potential and the LJ potential for liquid Ar.
The experimental results (open squares) are taken from ref.\onlinecite{Roder1987}.}
\end{figure}

As a first step, we have focused on the case of the system simulated with a WCA interaction potential using $n$~=~6,
in order to compare it to the corresponding LJ liquid. 
Figure~\ref{WCA} shows a plot of  the thermal conductivities calculated using the WCA potential 
vs. those using the LJ potential for a series temperatures and  packing fraction spanning a wide range,
i.e. from 100~K to
150~K  and from
0.240 to 0.427, respectively.
It is immediately seen that both the WCA thermal conductivities are 
identical (within the error bar of the calculation) to the LJ ones 
for the whole range of conditions explored. 
This shows that the attractive part of the LJ potential does not play a direct role
in the thermal conduction in such simple fluids.
Its impact is indirect only since it fixes the density of the system at a given thermodynamic point. 
It is worth noting from Figure~\ref{WCA} that the experimental thermal conductivities~\cite{Roder1987}
are also very well reproduced under corresponding conditions (i.e. the temperature variation at fixed $\eta$). 

\begin{figure}[t]
\includegraphics[width=\columnwidth]{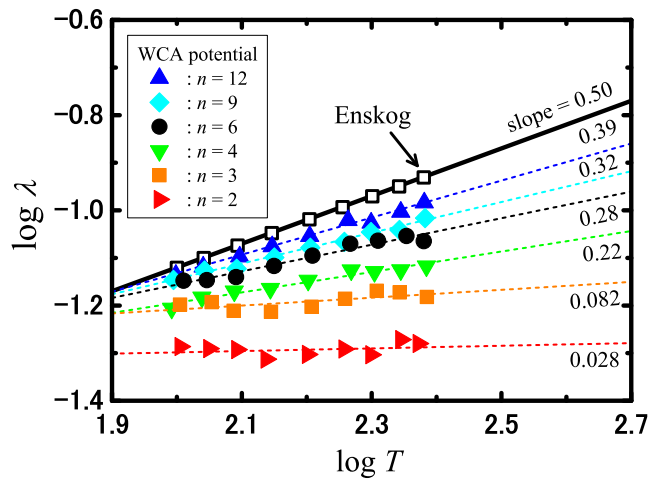}
\caption{\label{Tdep} Log-log plot of the temperature dependence of the thermal conductivity of
WCA fluids of varying softness $n = 2, 3, 4, 6, 9$, and 12, compared to the Enskog equation for hard spheres.
The linear fits indicate a temperature dependence in $T^x$, where $x$ is given by the slope of the line.}
\end{figure}

Figure~\ref{Tdep} shows the temperature dependence of the thermal conductivity 
with WCA interaction potentials of various softness ($n = 2, 3, 4, 6, 9,$ and 12)

, where $T$ ranges from 100~K to 250~K (at fixed $\eta$~=~0.351).
The results are also compared to the case of the hard sphere limit, 
which is provided by the Enskog equation~\cite{Heyes2001}
     \begin{eqnarray}
        \lambda_{\rm E} = 4 {\eta_{\rm HS}} \left[ \dfrac{1.02513}{y} + 1.23016 + 0.776483 y \right] {\lambda^{\rm B}}, \\
         y \equiv \dfrac{PV}{Nk_BT}-1 \approx 4{\eta_{\rm HS}} \left( 1 - {\dfrac{1}{2}} {\eta_{\rm HS}} \right) /
                      \left( 1-{\eta_{\rm HS}} \right) ^{3}, \label{eq:carnahan}
     \end{eqnarray}
where $\lambda^{\rm B}$ is the Boltzmann equation for a dilute gas,
     \begin{equation}
        {\lambda^{\rm B}} = \left( \dfrac{75k_B}{64{\sigma_{\rm HS}}^{2}} \right)
                             \left( \dfrac{k_BT}{m{\pi}} \right) ^{1/2},
     \end{equation}
and the second term in Eq.~\ref{eq:carnahan} is the Carnahan-Starling approximation for pressure, $P$.\cite{Heyes2001,Carnahan1969}
In a hard sphere fluid, 
following the Enskog equation, the thermal conductivity shows a temperature dependence of \Tdep. 
In the case of the WCA fluids, 
a temperature dependence of $T^{x}$ is also observed, in which the value of $x$ decreases when the potential becomes softer.
This progressive decoupling of the thermal conductivity with temperature indicates that the main mechanism at the origin of heat conduction in a fluid of soft spheres is no longer
due to the transport of
heat (energy) 
via the displacement of particles, as is the case for hard spheres%
, but rather to the ``collisional'' transport via interparticle
forces derived from the potential.~\cite{Hansen}
This interpretation also leads to the simple idea that a soft repulsion will damp the impact of collision between particles,
thus resulting in a relative decrease of the thermal conductivity.
In the case of LJ liquids like argon, the relative softness of the repulsive part of the potential is therefore the origin of the
\THdep
dependence.

\begin{figure}[t]
\includegraphics[width=\columnwidth]{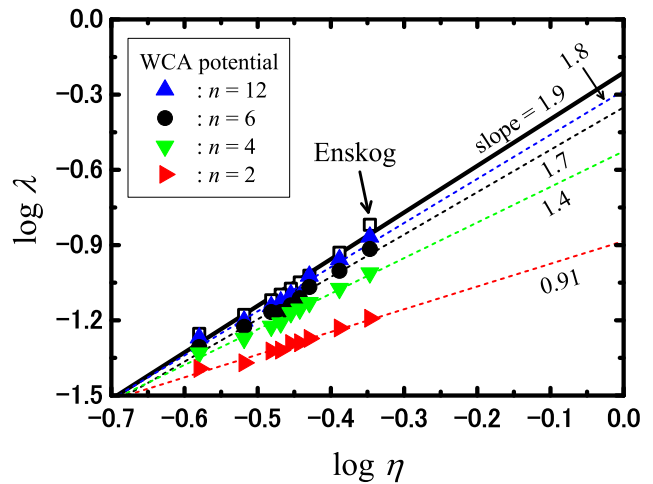}
\caption{\label{PFdep} Log-log plot of the packing fraction dependence of the thermal conductivity of  WCA fluids of varying softness $n = 2, 4, 6$, and 12, compared to the Enskog equation for diluted hard spheres. The linear fits indicate a packing fraction dependence in $\eta^y$, where $y$ is given by the slope of the line.}
\end{figure}

A similar approach is used for studying the packing fraction dependence%
, where $\eta$ is progressively switched from 0.263 to 0.450 (at fixed $T$~=~124~K) by varying the scaling parameter $a$ from 1.10 to 0.92 ($a$~=~1 corresponds to liquid argon).
As shown on Figure \ref{PFdep},the packing fraction dependence also becomes weaker in descending order of $n$.
The variation is less marked than for the temperature case, but again we can conclude that the \pfdep dependence which is observed
in the simulations of LJ fluids is due to the relative softness of the repulsive part of the potential.

\begin{figure}[t]
\includegraphics[width=\columnwidth]{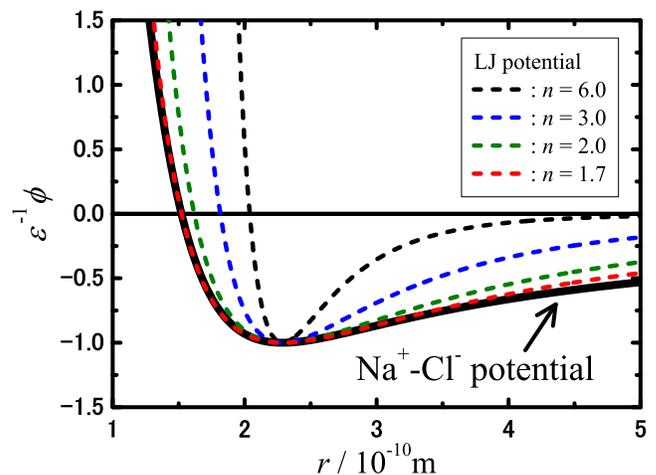}
\caption{\label{NaCl} Comparison of potential profile between the LJ potentials
with various power, $n$, and the Fumi-Tosi potential for NaCl. Each potential is normalized by
minimum of $\phi_{\rm LJ}$ or $\phi_{\rm Na^{+}-Cl^{-}}$, respectively.}
\end{figure}

It is then tempting, in the light of these findings, to try to assess the case of more complex fluids, such as molten salts. 
Here we will focus on the case of NaCl, 
which is well reprensented by the pair potential from Fumi and Tosi.~\cite{Fumi1964,Tosi1964}
Indeed, most of the transport properties of this salt, including the thermal conductivity are fairly well
reproduced with this potential even though polarization effects are not included.~\cite{Salanne2011:2}
Such a liquid now includes two components, and three different interaction potentials (Na-Cl, Na-Na and Cl-Cl)
should in principle be considered.
Nevertheless, due to the strong Coulomb ordering, contact occurs only between cations and anions,
so that we can restrict our attention to the Na-Cl one.
Figure \ref{NaCl} compares the Na-Cl Fumi-Tosi potential shape with that of a series of LJ-type fluids (i.e. with varying $n$)
for which $\epsilon$ is set to the value of the minimum (located at $r=r_{\rm min}$ of the Fumi-Tosi potential and $\sigma=r_{\rm min}/2^{1/n}$.
It is immediately seen that the softness of the Fumi-Tosi potential is very close to that of a LJ potential for which $n$ = 2
(by allowing non-integer values for $n$ and fitting the Na-Cl potential, we obtain a perfect matching for $n = 1.7$).
If the situation described before is still accurate, i.e. if the softness of the potential entirely determines the temperature
and fraction dependence of the potential, we can thus expect an almost null temperature dependence (at constant densities and packing fraction)
and a packing fraction dependence lower than 0.91 based on Figures \ref{Tdep} and \ref{PFdep}.
This is indeed the case, since a scaling law in \msnvpffTdep has been established in our previous study,
based on simulations across a wide range of thermodynamic states.~\cite{Ohtori2014,ohtori2009}

\section{\label{sec:level4} CONCLUSIONS}
Many properties of a liquid, including its structure and thermodynamics, depend on a subtle balance between the repulsive and
attractive forces due to the interactions between the atoms. In particular, its density will largely depend on the 
attractive ones. By comparing the thermal conductivities obtained from molecular dynamics simulations using Lennard-Jones 
potentials with those obtained with the repulsive-only Weeks-Chandler-Andersen potential, we have shown that, as soon as the 
density is fixed, the repulsion plays the main role in the heat conduction mechanism of simple fluids such as argon. By 
investigating WCA fluids with potentials of varying softness, we have then been able to explain the origin of the different packing fraction 
and temperature dependence of these liquids with the one of molten salts. These new findings will allow us to rationalize and
propose predictions of the thermal conductivities of many fluids, which is of high interest in many technical applications 
in which they are used for heat transfer purposes.

\section*{ACKNOWLEDGMENTS}
This work was partially supported by Grant-in-Aid for Scientific Research (c)
(Grants No. 21540382 and No. 24540397) from the MEXT-Japan.  MS would like to thank the Japan Society for the Promotion of Science for a short-term invitation fellowship which enabled this collaboration.

\nocite{*}


\providecommand{\noopsort}[1]{}\providecommand{\singleletter}[1]{#1}%
%


\end{document}